\documentclass[conference]{IEEEtran}
\IEEEoverridecommandlockouts
\usepackage{cite}
\usepackage{amsmath,amssymb,amsfonts}
\usepackage{algorithmic}
\usepackage{graphicx}
\usepackage{textcomp}
\usepackage{xcolor}
\usepackage{datetime}
\usepackage[numbers]{natbib}
\usepackage[shortlabels]{enumitem}
\newcommand{\mech}{\text{\bf \textrm{SkillCheck}}}
\newcommand{\trupeqa}{\text{\bf \textrm{TRUPEQA}}}

\usepackage{cleveref}
\crefname{claim}{claim}{claims}
\crefname{fact}{fact}{facts}
\crefname{algorithm}{algorithm}{algorithms}
\crefname{observation}{observation}{observations}
\crefname{equation}{equation}{equations}
\crefname{assumption}{assumption}{assumptions}
\crefname{hypothesis}{hypothesis}{hypotheses}

\newtheorem{theorem}{{\sc Theorem}}
\newtheorem{definition}{{\sc Definition}}

\def\BibTeX{{\rm B\kern-.05em{\sc i\kern-.025em b}\kern-.08em
    T\kern-.1667em\lower.7ex\hbox{E}\kern-.125emX}}
\begin{document}

\title{\mech: An Incentive-based Certification System using \emph{Blockchains}\\
\thanks{This work is supported by the IIT Kanpur fund number CS/2017198.}
}

\author{\IEEEauthorblockN{Jay Gupta and Swaprava Nath}
\IEEEauthorblockA{\textit{Indian Institute of Technology Kanpur}\\
\{jaygpt, swaprava\}@iitk.ac.in}
}
\maketitle

\begin{abstract}
Skill verification is a central problem in workforce hiring. Companies and academia often face the difficulty of ascertaining the skills of an applicant since the certifications of the skills claimed by a candidate are generally not immediately verifiable and costly to test. Blockchains have been proposed in the literature for skill verification and tamper-proof information storage in a decentralized manner. However, most of these approaches deal with storing the certificates issued by traditional universities on the blockchain. Among the few techniques that consider the certification procedure itself, questions like (a)~scalability with limited staff, (b)~uniformity of grades over multiple evaluators, or (c)~honest effort extraction from the evaluators are usually not addressed. We propose a blockchain-based platform named \mech, which considers the questions above, and ensure several desirable properties. The platform incentivizes effort in grading via payments with tokens which it generates from the payments of the users of the platform, e.g., the recruiters and test takers. We provide a detailed description of the design of the platform along with the provable properties of the algorithm.
\end{abstract}


\section{Introduction}
\label{sec:intro}

Blockchain technology has seen an unprecedented growth after being introduced in 2008. First used as a peer-to-peer ledger for registering the transactions, it has seen its own technological development as well as its adoption in various applications. It is desirable in many applications primarily because it eliminates any third-party intermediary and allows users to make their transactions directly. In a blockchain, each node in a decentralized network of peer nodes (1)~holds a replica of a common ledger, (2)~writes an entry to its own ledger when it has agreement from the other nodes in the network, (3)~broadcasts any transaction made by its user to the other nodes in the network, and (4)~checks, on a regular basis, that the ledger it holds is identical to the ones across the network \cite{alammary2019blockchain}.

On another hand, the rapid development of the online education, powered by the massive open online courses (MOOC), has also seen an unprecedented growth. In the process of this growth, it needed some of the properties that are ubiquitous in the blockchains, most notably, (i)~the absence of a central authority and (ii)~verifiable certifications. Skill certification today demands need-based skills to be certified which is a little difficult to come by in a standard university system. For instance, a course on android app development may not find a place in a university curricula, while a student skilled in that may be indispensable for certain organizations. Yet, the organization has to hire the student uncertified or has to test the skill by itself, which is expensive. 

To bridge this gap and to provide skill certification on demand, the need for blockchains in education has emerged \cite{williams2019does}. This need is also reflected in the European commission policy report \cite{grech2017blockchain}, which agrees that the blockchain applications in education is still at infancy, and recommends that
\begin{quote}
To ensure development of open blockchain implementations we recommend that the EU
in collaboration with Member States consider creating and promoting a label for `open'
educational records, which enshrines the principles of recipient ownership, vendor
independence, and decentralised verification.
\end{quote}
The literature on application of blockchains in education is heavy with the management of certificates issued by the standard authorities. A survey by \citet{alammary2019blockchain} shows that almost 41\% of the systematically surveyed articles dealt with this issue. However, a major impediment to the skill certification is to get the right instructor and evaluators that can certify the skills of candidates. This is a challenging task since there is little work in this domain that study the {\em quality} of the certificates or on ensuring the {\em best effort} of the evaluators. In this paper, we take an {\em incentive}-based approach for the different players in the education blockchain network and investigate what guarantees are possible to provide in the {\em quality of certifications}.

\subsection{Our contributions}
\label{sec:contribution}

We propose a blockchain-based education platform called \mech, which handles large classes with limited number of teaching staff, provides uniformity of grades, and ensures honest effort from the evaluators. This platform can be implemented inside a university or an organization, or can be implemented by onboarding multiple universities or organizations on a single large blockchain. It allows any user to offer a course according to the need-based skill certifications of the other users. Hence, it has the flexibility of designing and certifying {\em non-traditional} courses in an {\em asynchronous} manner, i.e., the beginning or end does not need to align with any semester/quarter. The instructor can choose a set of evaluators in the blockchain network who are skilled in that topic, e.g., individuals who have been certified in the same system earlier. \mech\ uses crypo-tokens for transferring credit for {\em good} evaluation to the evaluators (\Cref{sec:workflow}). The underlying mechanism of \mech\ ensures that the rewards of evaluation is designed such that the individual biases of evaluators have no effect in the final score received by the candidate. Therefore, it is ensuring {\em uniform} grading of the papers. Also, the mechanism ensures that if an evaluator exerts effort to reduce her {\em noise} of grading, her pay-off increases (\Cref{thm:bi-rm}). We provide layer-wise divisions of the functionalities of the platform (\Cref{sec:layer}) and a detailed description of the implementation of the platform (\Cref{sec:implementation}) so that it is reproducible.

The complete details are available in the full version of the paper \cite{Gupta2020}.

\subsection{Related work}
\label{sec:related}

The nascent research area of blockchains in education can be classified in many strands. Here we will discuss about two largest strands. The first strand is focused towards certificates management \cite{srivastava2018distributed, nespor2019cyber, funk2018blockchain, lizcano2019blockchain, sharples2016blockchain, gresch2018proposal}, which is a secure decentralized way to store the traditional university certificates. Our work is closest to the second strand of literature that deals with the management of skill or competencies \cite{farah2018blueprint, williams2019does, duan2017education, zhao2019design, srivastava2018distributed, lizcano2019blockchain, mikroyannidis2018smart}. These works focus on generating the competencies based on the evaluations of the evaluators. However, the evaluators are assumed to be honest graders and there is no normalization of the scores given by different evaluators, if used. 
\citet{alammary2019blockchain} provide a nice classification of the literature based on their category of applications.
The literature, however, is thin on the algorithmic guarantees on competencies of the evaluators. Our approach fills this important gap with a platform with provable algorithmic guarantees for skill certification.

\section{Design Workflow}
\label{sec:workflow}

The basic platform of skill certification in this paper is a blockchain with users who can have exactly one of the {\em four} roles at any given instant: (a)~the instructor, who designs and conducts an exam, sets the guidelines of grading, and also partially grades answerscripts, (b)~the evaluators, who are individuals with sufficient expertise in the topics of the exam -- they can be past creditors of the course, (c)~the candidates seeking certification, (d)~the viewers of the certificates, e.g., the recruiters who are looking for individuals of a certain skill. We assume that the number of candidates is quite large for the instructor to check. Therefore, he needs evaluators, e.g., the teaching assistants, who help him grading the papers. However, the evaluators may spend little or no effort in the grading carefully, which can lead to a bad quality of certification. So, in such a context of decentralized evaluation and certification, we need certain desirable properties of any platform, given as follows.
\begin{itemize}[noitemsep,leftmargin=*,topsep=0pt,parsep=0pt,partopsep=0pt]
 \item {\em Scalability}: the platform should handle a large number of candidates.
 \item {\em Uniformity of scores}: if the scores are given by different evaluators having different degrees of noise, it is necessary that the final grade is normalized in some way to ensure uniformity of the grades.
 \item {\em Honest effort extraction}: evaluators must be incentivized to maximize their effort in grading the papers.
\end{itemize}
The certification obtained via this platform will always be stored in a blockchain. Therefore, the properties like verifiability, absence of a central repository of certificates, and authenticity of the records are inherited by default, and are not mentioned as separate desiderata.

To ensure the desirable properties, our approach is to exploit the fact that a grading mechanism (a)~can have some instructor-graded answer-scripts (which we call {\em probes}) to measure the {\em quality} of the evaluators, and (b)~allow students to raise a {\em regrading request} for incorrect grades which can be corrected by the instructor. Hence, eventually all `true' grades of the papers, where the `given' scores are below the `true' scores, will be obtained, and this information will be used to deter the evaluators from deliberately under-performing. Our approach is inspired by the \trupeqa\ mechanism \cite{chakraborty2018ensuring}, but modified for our setting.


In the following, we outline the functional structures of our proposed platform \mech. We will discuss the model of the evaluation process and the provable properties it satisfies in subsequent sections.

\smallskip \noindent
\textit{\mech\ token. }
Crypto tokens are well-known holdings in the crypto space and are usually called {\em cryptocurrency} in the blockchain parlance. It is a standard currency which is used to make and receive payments on a blockchain network. In contrast to fiat currency, crypto tokens are a special kind of virtual representation of a particular asset or utility, that reside on the blockchain network.

In our setting, all four types of users of \mech\ will have individual {\em wallets} attached to their account, which contain the amount \mech\ tokens that user (represented by a node in the network) hold, and individual {\em portfolios} showing their skill score and evaluation scores (relevant for the candidates and evaluators) that constitute their certificates. The skill score shows the knowledge of the candidates on a subject and evaluation score reflects the evaluator's skill in grading an answersheet. To begin with, the wallets can be recharged by converting fiat currency into the tokens. However, we propose to waive the transaction costs of the evaluators, since they are the workforce for the \mech\ platform. 

\begin{figure*}[ht]
    \centering
    \includegraphics[width=0.7\linewidth]{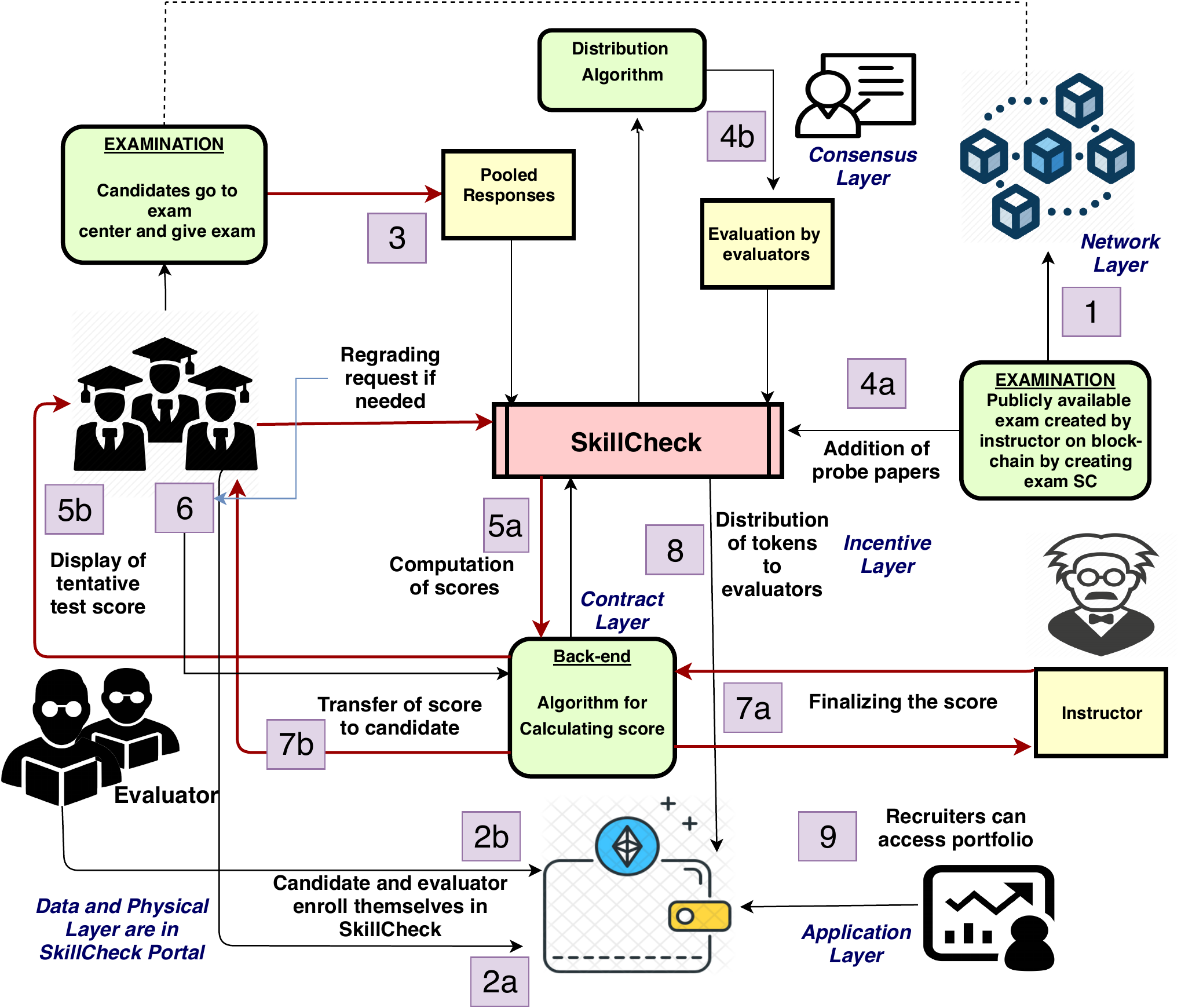}
    \caption{The workflow diagram in \mech.}
    \label{fig:workflow}
\end{figure*}

The examiners and candidates pay \mech\ tokens as a fee to conduct any examination and to get certified respectively. The evaluators also need to pay a small threshold token to enroll themselves as an evaluator. 
The total collected token (from initial token charged to the examiner and small token paid by the evaluators) will be distributed to evaluators based on the evaluation score that evaluator will earn by grading the papers. The steps of the workflow of \mech\ is given as follows (see \Cref{fig:workflow}).

In step 1, an examiner submits an exam on blockchain network by calling a public function to create a new copy of a {\em Smart Contract}\footnote{A smart contract, also known as a cryptocontract , is a computer program that directly controls the transfer of digital currencies or assets between parties under certain conditions. An algorithm can be encoded into a smart contract which automates the payment process.} (SC). Placing an examination on the blockchain costs him some fixed number of tokens. Afterwards, the paid token for the submission is locked within the SC. 
Users' wallets associated with the blockchain network contains the portfolio in which users will have skill and evaluation scores, along with \mech\ token which can be earned by evaluating papers or purchasing from the system. 
Once the question paper of the exam is on the blockchain network, eligible candidates can enroll themselves in examination by calling the public function of the SC. The students have to go to the local examination centre for giving the exam.

In step 2, candidates and evaluators enroll themselves using their public keys.

In step 3, the examination will be conducted at a test center to prevent identity theft. Every candidate gets a fixed number of maximum attempts to get themselves certified. The candidates need to solve the paper and submit the responses by calling the private function of the SC, which ensures the prevention of data leakage threat. The SC will contain all pooled answers of candidates. 
    
In step 4, the examiner checks a constant number of answerscripts that constitutes the probe papers. This is needed as our algorithm (explained in \Cref{sec:contract}) needs the probes to estimate the accuracies of the evaluators. The scores of the probe papers are assumed to be benchmarks as it is graded by the instructor.
A distribution strategy is implemented in the SC that distribute sets of papers to all evaluators. Every evaluation gets an equal mix of anonymous probe and non-probe papers.  The evaluators check the papers assigned to them and submit their scores for the assigned papers.  A small gas price for this evaluation transaction is charged to the evaluators. This price is a constant and is later reimbursed by the platform. The evaluators need to call the SC function and submit evaluated marks to contract.

In step 5, our algorithm computes the skill scores of the candidates participating in the exam. The scores are shared with the candidates.

In step 6, the candidates can raise a regrading request if they feel the scores are unfair. The paper is then checked by the instructor to find the ``true'' score. Our algorithm keeps a penalty for frivolous regrading requests to entertain only reasonable regrading requests. Due to the properties of our algorithm, which we prove later, the number of regrading requests are expected to be small.

In step 7, based on the evaluations of the instructor on the papers that went through the regrading process and the papers for which there were not any regrading request, the final skill scores of the candidates and the evaluation scores of the evaluators are updated in the SC. For the papers that did not raise any regrading request, the final score is the score given by the algorithm in step 5.

Step 8 computes an {\em appropriate} distribution of collected token from evaluators and examiner and the SC will transfer the collected tokens to evaluators. The skill scores are registered in the profile of the candidates via the SC.

In step 9, the viewers, e.g., the recruiters, can view the skill and evaluation scores of the users by paying tokens.


\section{Layer-wise Design of \mech}
\label{sec:layer}

\begin{figure}[ht]
\centering
    \includegraphics[width=0.7\linewidth]{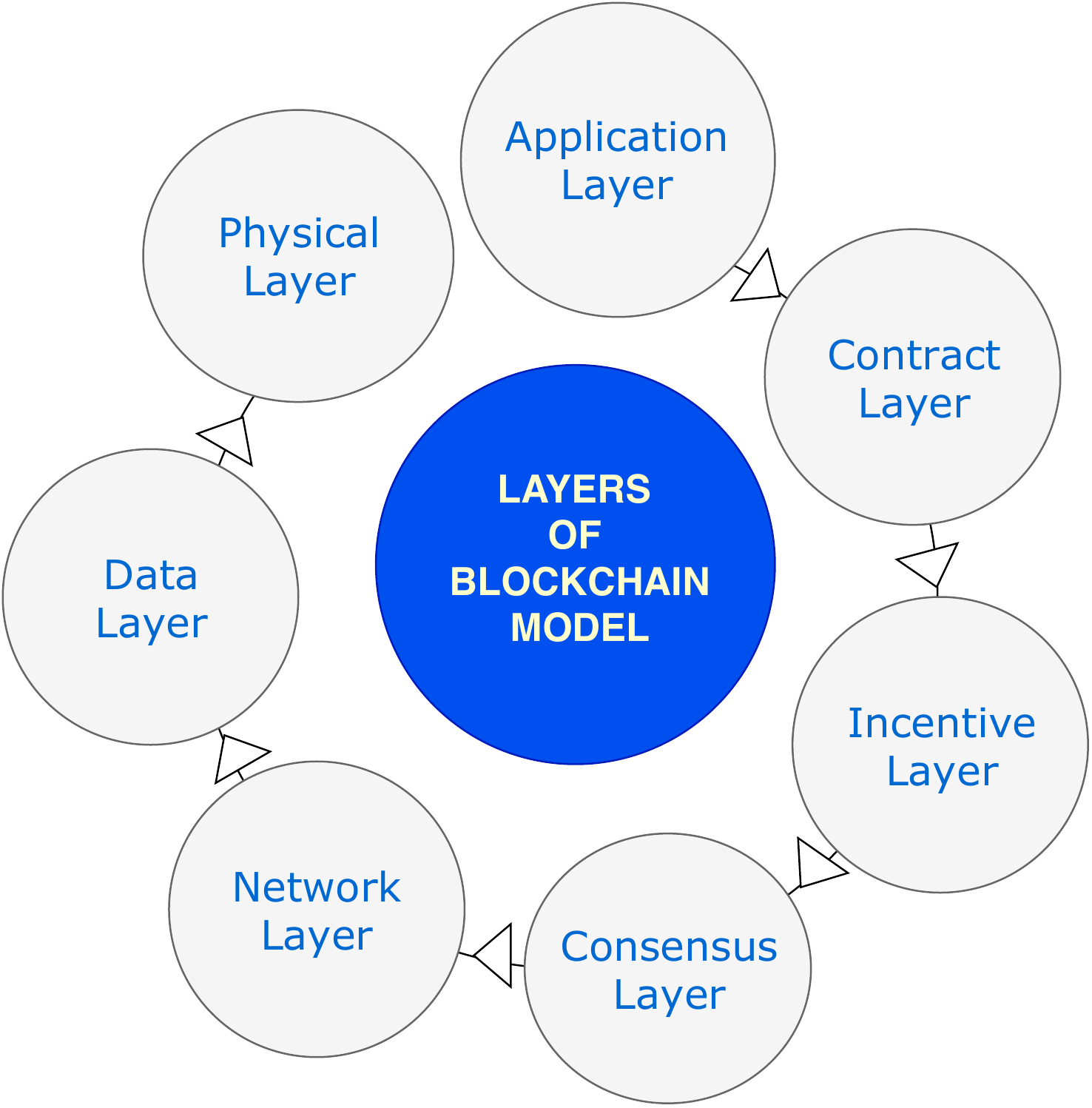}
    \caption{Layered architecture of the \mech\ platform.}
    \label{fig:layer}
\end{figure}

Since the development of the platform \mech\ needs all functionalities of a blockchain environment to be implemented, it is often easier to break them down into multiple layers of specific functionalities. We follow the layered structure given by \citet{yuan2016towards}. 
In this section, we provide a detailed view of the first three layers, i.e., the  application, contract, and incentive layers. For the rest of the layers, i.e., the consensus, network, data, and physical layers, we use the standard implementation of the Ethereum framework and its default configuration. \Cref{fig:layer} shows the layered architecture of our proposed platform.

\subsection{Application Layer}

This layer deals with the application of skill verification of users, authentication of certificates issued by the various institutes using blockchain networking, and stores the tokens in the user wallets. In the \mech\ platform, this layer deals with the development of the user interface that shows their profile with (a)~the certificates issued by the platform, (b)~the coins in the wallet, and (c)~the roles that they can take--e.g., instructor, candidate, evaluator, or recruiter.

\subsection{Contract Layer}
\label{sec:contract}

This layer mainly consists of various scripts for functionalities, which serve as a necessary means to communicate with the whole network, including blocks stored in the blockchain. This layer also implements the central algorithm of \mech. The algorithm and functionalities are encoded into a smart contract, which is a group of self-authenticating, self-administering, and self-driving quick response rule system that is stored on the blockchain. To present the central algorithm of \mech, we develop the basic notation. The notation and the desiderata are similar to the properties mentioned in \cite{chakraborty2018ensuring}. However, unlike \cite{chakraborty2018ensuring}, here the evaluation of the papers are done by evaluators who are different from the test takers.

\subsubsection{Setup}

Each candidate interested in getting certified has written an exam that needs to be evaluated. Let $N=\{1,\ldots,n\}$ represent the set of the candidates and $M=\{1,\ldots,m\}$ denote the set of evaluators. We use $i$ as the index for an evaluator and $j$ as the index for a paper (of a candidate). Out of the $n$ total papers,  $\ell (<< n)$ papers are evaluated by the instructor and these papers are called the {\em probe} papers. We assume that through this grading process, the {\em true grades} of these probe papers are known to the designer (e.g., the instructor of the course).
The scores belong to a closed interval $S \subset \mathbb{R}$. 

An {\em evaluation mechanism} $E$ is the tuple $\langle G, \mathbf{r}, \mathbf{t} \rangle$, where 
\begin{itemize}[noitemsep,leftmargin=*,topsep=0pt,parsep=0pt,partopsep=0pt]
 \item $G$ is an assignment function $G: M \mapsto 2^N$ that maps papers to evaluators. 
 The set of papers graded by $i$ is $G^{-1}(i) \triangleq \{j \in N : i \in G(j)\}$.
 \item $\mathbf{r} : \times_{i\in N} S^{G(i)} \to S^n$, where the $j$th component $r_j (\cdot)$ is the function assigning the skill score of candidate $j$ based on the scores reported by the assigned evaluators.
 \item $\mathbf{t} : \times_{i\in N} S^{G(i)} \to S^n$, where $t_i (\cdot)$ is the function that yields the evaluation score to evaluator $i$.
\end{itemize}
The set $P_i \subset G^{-1}(i) $ and $NP_i = G^{-1}(i) \setminus P_i$ denotes respectively the probe and non-probe  papers assigned to $i$.  We define the co-evaluators of individual $i$ as $CG_i = \cup_{j \in NP_i} G(j) \setminus \{i\}$, i.e., the set of evaluators who grade at least one common non-probe paper with $i$.

\subsubsection{Model of true and reported scores} 
\label{sec:model}

We assume that the generation of the true scores and the error model of the evaluators is given by the $\mathbf{PG}_1$ model of evaluator bias and reliability as described in \citet{piech2013tuned}, described as follows.
 \begin{itemize}[noitemsep,leftmargin=*,topsep=0pt,parsep=0pt,partopsep=0pt]
  \item The true score $y_j$ for paper $j$ is distributed as $\mathcal{N}(\mu,1/\gamma)$, for all $j \in N$. The parameters of this distribution are learned from the historical data of examinations. 
 \item The reported score of paper $j$ by evaluator $i$, given by $\tilde{y}_{j}^{(i)}$, is distributed as $\mathcal{N}(y_{j}+b_{i},1/\tau_{i})$, where $b_i$ and $\tau_i$ are called the {\em bias} and {\em reliability} of $i$ respectively. 
 \end{itemize}
  
The bias term may be interpreted as either strategic (manipulating grades in a particular direction consciously) or non-strategic (subconsciously generous or strict evaluators). We assume that an evaluator can {\em choose} her reliability in the grading process. Intuitively, bias resembles to what extent the evaluator is {\em shifting the mean} of the score and reliability denotes how much {\em effort} she is putting in to find the score accurately.
A crucial assumption here is that an evaluator grades all papers in an exam (particularly, the probes and non-probes) with the same bias and reliability. We believe that this assumption is not too restrictive in an environment where the evaluators cannot distinguish the probes from the non-probes (as they are given a mix of them without mentioning the distinction).
We use the shorthand $\theta_i := (b_i, \tau_i) \in \mathbb{R} \times \mathbb{R}_{\geqslant 0}$, where $b_i$ (potentially) and $\tau_i$ are evaluator $i$'s strategic choices.

\subsubsection{Description of our mechanism $\langle G^*, \mathbf{r^*}, \mathbf{t^*} \rangle$}

The components of our evaluation mechanism $\langle G^*, \mathbf{r^*}, \mathbf{t^*} \rangle$ are as follows. Let $K$ be an {\em even} predetermined number of papers each evaluator grades. This number is an exogenous design choice of the platform.
\paragraph{Paper assignment rule $G^*(\cdot)$} 
We impose (a)~$G^*(j) \neq \emptyset, \ \forall j \in N$, (b)~$|P_i| = K/2, \ \forall i \in M$, and (c)~$|NP_i| = K/2, \ \forall i \in M$, i.e., every paper is graded by at least one evaluator, and every evaluator grades $K/2$ probes and $K/2$ non-probe papers. The evaluators know the numbers of probe and non-probe papers assigned to them, but cannot tell them apart.

\paragraph{Skill and evaluation scores} Once the evaluators scores are collected, our mechanism statistically estimates the error parameters $\hat{\theta}_i : S^{|P_i|\times |P_i|} \mapsto \mathbb{R} \times \mathbb{R}_{\geqslant 0}$ of each evaluator $i$, from the comparison between reported grades $ \tilde{y}_{j}^{(i)}$s and true grades $y_j$s on the probe papers $P_i$. In our particular case, this is given by  $\hat{\theta}_i = (\hat{b}_i, \hat{\tau}_i)$, and $\hat{b}_i = \frac {\sum_{j \in P_i} \tilde{y}_{j}^{(i)} - {y}_{j} }{|P_i|}$ and $\hat{\tau}_i = \frac{|P_i|-1}{\sum_{j \in P_i} (\tilde{y}_{j}^{(i)}-(y_{j}+\hat b_{i}))^2}$. 
Below we describe how the estimated parameters are used in generating the skill scores of the candidates and the evaluation scores to evaluators.

\begin{definition}[Score computing mechanism]
 \label{def:score-welfare}
 We define a specific score computing rule that would be used in aggregating the evaluator-reported-scores into a final score on each exam. We also define a social welfare criteria that would be used to describe the evaluation scores.
 \begin{itemize}[noitemsep,leftmargin=*,topsep=0pt,parsep=0pt,partopsep=0pt]
  \item The score computing function $\mathbf{r}^* = (r_j^* : j \in N)$ of an evaluation mechanism $E$ is {\bf inverse standard-deviation weighted de-biased mean (ISWDM)} if it assigns the instructor-verified grade on every probe paper. For every non-probe paper $j$, it assigns
  \[r_{j}^{*}(\mathbf{\tilde{y}}^{G(j)}_{j}, \hat{\boldsymbol{\theta}}_{G(j)}) = \frac{\sqrt{\gamma}\mu+\sum_{i \in G(j)}\sqrt{\hat{\tau}_{i}}(\tilde{y}_{j}^{(i)}-\hat{b}_{i})}{\sqrt{\gamma}+\sum_{i\in G(j)}\sqrt{\hat{\tau}_{i}}}.\] where $\tilde{y}_{j}^{(i)}$ is the score assigned by the $i$th evaluator and $(\hat{b}_{i},\hat{\tau}_{i})$ are the parameters we estimated for her.
  \item The scores $\mathbf{r}^*$ are released to the candidates, and they can raise a regrading request if they feel the score is unfair. The instructor checks the paper in such a case and decides the final score $y_j$, which we assume to be the true score of the paper. If candidate $j$ do not raise a regrading request, the mechanism assumes the true score to be $r_j^*$. Here we can explicitly say that the total effort(in checking all papers himself) made by the instructor, has been reduced by a substantial amount.
\footnote{To avoid the case where every candidate may raise a regrading request (because it does not hurt), we can impose penalties for frivolous requests. This threat mechanism works well in practice and has been part of various online grading tools.}
  \item The {\bf social welfare} at a score $r_{j}^{*}$ for paper $j$ when the true score is $y_j$ is denoted by
  \[W_j^{*}(\mathbf{\tilde{y}}^{G(j)}_{j}, \hat{\boldsymbol{\theta}}_{G(j)}, y_j) = R(r_j^*(\mathbf{\tilde{y}}^{G(j)}_{j}, \hat{\boldsymbol{\theta}}_{G(j)}), y_j).\]
where $\mathbf{\tilde{y}}^{G(j)}_{j}$ is the vector of evaluated scores reported on paper $j$, and $\hat{\boldsymbol{\theta}}_{G(j)}$ is the vector of estimated error-parameters for the relevant evaluators $G(j)$.

We assume that $R(x,x)=0\geq R(x,y)=R(y,x)$ for all $x,y \in S$, and $R(x_1,y_1) \geqslant R(x_2,y_2)$ if $|x_1-y_1| < |x_2-y_2|$.
One example of such a function would be $R(x,y)=-(x-y)^2$, which calculates the squared error in assigned scores. 
  \item The {\bf social welfare} at a score $r_{j}^{*}$ for paper $j$ {\bf without evaluator $i$} when the true score is $y_j$ is denoted by $W^{(-i)*}_j=W_j^*(\mathbf{\tilde{y}}^{G(j) \setminus \{i\}}_{j}, \hat{\boldsymbol{\theta}}_{G(j) \setminus \{i\}}, y_j)$
  where $W^*_j(\cdot)$ is defined as before.
  \item The evaluation score of evaluator $i$ for grading paper $j \in NP_i$ is given by $t_i^{j,*} = \alpha(W_j^{*} - W_j^{(-i)*})$, where $\alpha>0$ is a constant chosen at the designer's discretion. The total transfer to evaluator $i$ is therefore $t_i^* = \sum_{j \in NP_i} t_i^{j,*}$.
 \end{itemize}
 The parameters $\gamma, \mu, b_i$, and $\tau_i$ are from the $\mathbf{PG}_1$ model of \citet{piech2013tuned} as defined at the end of \Cref{sec:model}.
\end{definition}
We will use the shorthands $W_j^{*}$ and $W_j^{(-i)*}$ for the above two expressions when the arguments of such functions are clear from the context.

\subsubsection{Incentives and design desiderata}
In this section, we describe the preferences of the population participating in the mechanism. A candidate $j$ can get only skill scores, $r_j^*$ or $y_j$. Her strategy to improve this score is via putting more effort in getting the answers correct and raise correct regrading requests. Hence there is no strategic aspect in a candidate strategy. An evaluator $i$, however, can have a conscious or unconscious bias $b_i$ and can choose the amount of effort, represented via the reliability $\tau_i$, that she can exert to grade the papers assigned to her. The pay-off (we use pay-off and utility interchangeably in this paper) of the evaluator is the evaluation score $t_i^*$. Depending on how $t_i^*$ is related to the bias and reliabilities, evaluator $i$ can choose those two parameters to maximize her incentive. This is where the incentive design question is important. Denote the utility of evaluator $i$ by $u_i^{(G^*, \mathbf{r^*}, \mathbf{t^*})} (b_i, \tau_i, b_{-i}, \tau_{-i}) = t_i^*$.~\footnote{The notation `$-i$' denotes all the evaluators except $i$.} 

The objective of the \mech\ platform is to ensure that the evaluation is done with uniformity of skill scores and by extracting honest effort from the evaluators. Hence the following properties are desirable in such a setting.

A few different uncertainties are resolved after the evaluator $i$ chooses her decision variables $(b_i,\tau_i)$, and before $\mathbf{r}^*$ and $\mathbf{t}^*$ are computed. These are: (1) the scores finally reported by this evaluator $i$, which on any paper $j$, follow the distribution $f(\tilde{y}_{j}^{(i)}|y_j) \sim \mathcal{N}(y_{j}+b'_{i},1/\tau_{i})$, (2) the decision variables $(b_k,\tau_k)$ chosen by every co-evaluator $k$, (3) the true score $y_j$ on paper $j$, (4) the scores finally reported by a co-evaluator $k$, which on any paper $j$ follow the distribution $f(\tilde{y}_{j}^{(k)}|y_j) \sim \mathcal{N}(y_{j}+b_{k},1/\tau_{k})$, conditional on the decision variables $(b_k,\tau_k)$. 
The properties defined below consider the evaluator $i$'s expected utility from the choice of strategies she makes, where the expectation is only being taken with respect to the distribution of $i$'s grade-evaluation process $f(\tilde{y}_{j}^{(i)}|y_j) \sim \mathcal{N}(y_{j}+b_{i},1/\tau_{i}), \forall j \in G^{-1}(i)$, and they hold for any realization of the other uncertainties (2) to (4).

\begin{definition}[Ex-Post Bias Insensitivity (EPBI)]
 \label{def:bi}
 An evaluation mechanism $E = \langle G, \mathbf{r}, \mathbf{t} \rangle$ is {\em ex-post bias insensitive}, if the expected utility of every evaluator $i \in M$, where expectation is taken with respect to the distribution of $i$'s grade-evaluation process, is independent of the bias $b_i$. This holds irrespective of the biases and reliabilities chosen by the other evaluators $k \neq i$, {\em and} the realized value of the true score $y_j$ and reported scores of the different evaluators. Mathematically, for all $\{\tilde{\mathbf{y}}^{(k)}, b_k, \tau_k\}_{k \neq i}$
 \begin{align}
  \label{eq:dsbi}
 \lefteqn{\mathbb{E}_{ \tilde{\mathbf{y}}_{G^{-1}(i)}^{(i)}|\mathbf{y}_{G^{-1}(i)} } u_i^{(G, \mathbf{r}, \mathbf{t})} (b_i, \tau_i, b_{-i}, \tau_{-i}) } \nonumber \\
 &=\mathbb{E}_{ \tilde{\mathbf{y}}_{G^{-1}(i)}^{(i)}|\mathbf{y}_{G^{-1}(i)} } u_i^{(G, \mathbf{r}, \mathbf{t})} (b'_i, \tau_i, b_{-i}, \tau_{-i}).
 \end{align}
\end{definition}
EPBI ensures that the expected utility of a evaluator is insensitive to her bias for (a)~every realization of the random variables governing the true and reported scores, (b)~the bias, reliabilities chosen by the other evaluators, and (c)~her own reliability. The next property discusses the dependency with reliability.

\begin{definition}[Ex-Post Reliability Monotonicity (EPRM)]
 \label{def:rm}
 An evaluation mechanism $E =\langle G, \mathbf{r}, \mathbf{t} \rangle$ is {\em ex-post reliability monotone} if for every evaluator $i$, the utility is monotonically non-decreasing with her reliability irrespective of the biases and reliabilities chosen by the other evaluators {\em and} the realizations of the true scores and the scores reported by the different evaluators. Mathematically, for all $\{\tilde{\mathbf{y}}^{(k)}, b_k, \tau_k\}_{k \neq i}$
 \begin{align}
  \label{eq:dsrm}
 \lefteqn{\mathbb{E}_{ \tilde{\mathbf{y}}_{G^{-1}(i)}^{(i)}|\mathbf{y}_{G^{-1}(i)} } u_i^{(G, \mathbf{r}, \mathbf{t})} (b_i, \tau_i, b_{-i}, \tau_{-i}) } \nonumber \\
 &\geqslant \mathbb{E}_{ \tilde{\mathbf{y}}_{G^{-1}(i)}^{(i)}|\mathbf{y}_{G^{-1}(i)} } u_i^{(G, \mathbf{r}, \mathbf{t})} (b_i, \tau_i', b_{-i}, \tau_{-i}), \forall \tau_i > \tau_i'.
 \end{align}
\end{definition}
Note that these properties are stronger than a similar definition in {\em dominant strategies}. Since there are random variables like the true scores $\mathbf{y}$ and the reported scores $\mathbf{\tilde{y}}$ respectively, and they can change the utility expression, a dominant strategy definition will only require the (in)equalities to be satisfied after taking {\em interim} expectation over some relevant distribution over those variables. However, the {\em ex-post} properties require them to be satisfied for {\em every} realization of these random variables.

Our central result on the proposed evaluation mechanism is as follows. We defer the proof to the appendix for a cleaner presentation.
\begin{theorem}[Bias and Reliability]
 \label{thm:bi-rm}
 Mechanism $(G^*, \mathbf{r^*}, \mathbf{t^*})$ is EPBI and EPRM.
\end{theorem}

This result ensures that any rational evaluator should not put any bias to the evaluations and work at their highest reliability. The $(G^*, \mathbf{r^*}, \mathbf{t^*})$ mechanism forms the backbone of \mech. Hence the platform satisfies the properties as that of this mechanism. In the next layer, we discuss how the incentives of the evaluators are implemented.


\subsection{Incentive Layer}
\label{sec:incentive}
Incentive layer is designed to incorporate the rewards to the participants (the evaluators in our case) to satisfy the objectives of the blockchain network. In \mech, the desirable properties are EPBI and EPRM. These are obtained through the award of the evaluation score $t_i$ to evaluator $i \in M$. Our platform transfers $t_i$ amount of tokens to evaluator $i$, which they can convert to fiat currency. However, we need to ensure that the platform does not run into an overall loss after such transfers. The platform earns the tokens from the payments of (a)~the candidates who are willing to get certified, (b)~the recruiters, who wants to view the certificates of the candidates, and (c)~the instructor, who needs to pay to offer a course on \mech. Let the collective amount from all the tokens received in these way is $K_\text{net}$. \mech\ ensures that the expected sum of transfers, i.e., $\mathbb{E}(\sum_{i \in M} t_i)$, is below $K_\text{net}$, where the expectation is taken over the prior distribution of the skill scores, evaluators' error distributions, and the observed scores of the evaluators. The distributions can be obtained from the historical data of the course. We call this property {\em expected budget balance}, which can be achieved by setting the parameter $\alpha$ in the expression of $t_i^j$ based on the $K_\text{net}$ and the expected value of the sum of the evaluation scores.

For a practical implementation, the estimate of $\alpha$ may be set conservatively, i.e., that ensures that the $K_\text{net}$ is less likely to get exhausted. However, the choice of $\alpha$ for a specific exam needs to be derived empirically using the available historical information of that exam. 

The rest of the layers are implemented according to any standard protocol of blockchain platforms used in practice. We follow the implementation of the Ethereum and the details are given in the next section.

\section{Blockchain Implementation}
\label{sec:implementation}

\begin{figure*}[t!]
    \centering
    \includegraphics[width=0.85\linewidth]{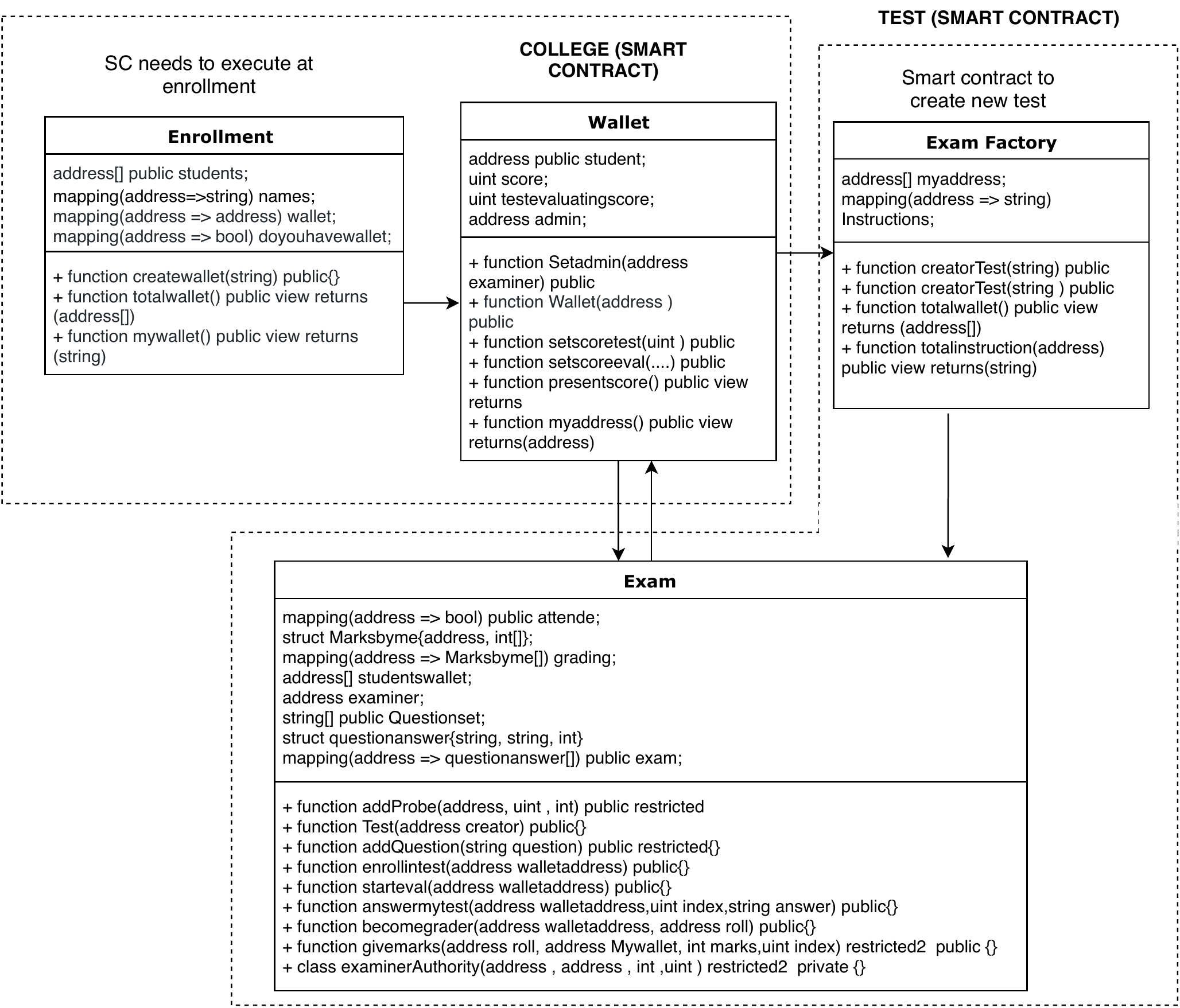}
    \caption{Overview of the smart contracts of \mech.}
    \label{fig:smartcontract}
\end{figure*}

To implement \mech, we use the Rinkeby Test network provided by the Ethereum Foundation. In the current implementation, we have used the Ethereum private network to establish a blockchain system. We have used Solidity to perform the primary functionality of a smart contract (SC). Every action in the platform \mech, e.g., checking the probe papers by the instructor, evaluations of the answerscripts by the evaluators, raising regrading requests by the candidates, and assignment of final skill and evaluation scores by the instructor, is considered a transaction, which is executed via a cryptographically signed contract that runs on the network. The codes for full implementation can be found at: \texttt{https://github.com/jaygpt/Skillcheck}.

The implementation of \mech\ is divided into {\em three} main segments: (1)~the {\em front-end} that deals with the user interface, (2)~the {\em smart contracts} where the algorithms are implemented, and (3)~the {\em back-end} which connects \mech\ to the Ethereum blockchain. These components are explained as follows.

\begin{figure}[ht]
    \includegraphics[width=\linewidth]{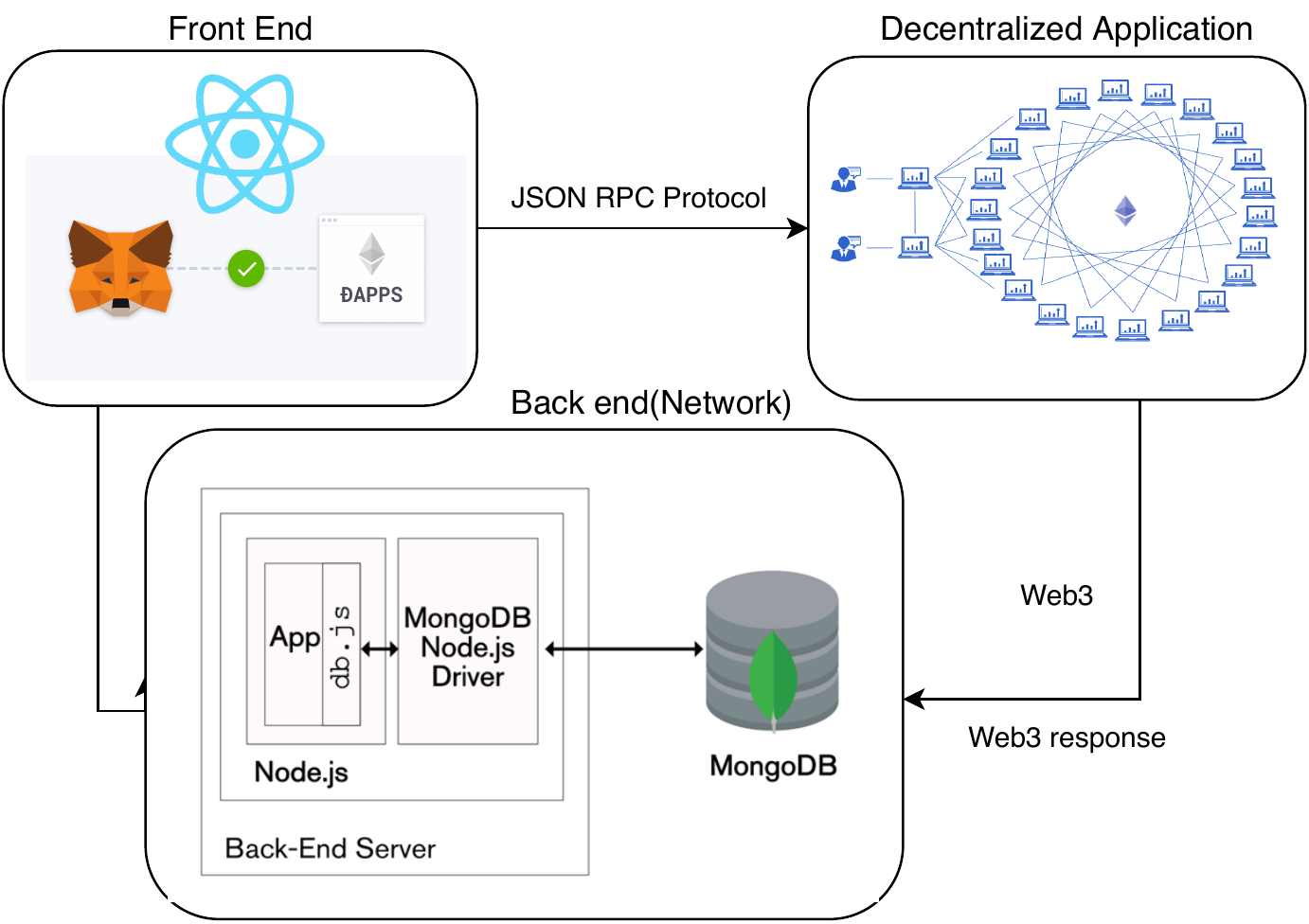}
    \caption{The architecture of \mech.}
    \label{fig:architecture}
\end{figure}

\subsection{Back-end}
Like most of the Ethereum's decentralized application (DApp), \mech\ also has Node.js as its backend framework. 
We have used a combination of Node.js and MongoDB to maintain the database and to manage the connection between the database and the blockchain interactions and to display using our frontend and render methods. However, \mech\ uses web3.js JavaScript library which allows the client-side to interact with the blockchain including a local or a remote Ethereum node, using a HTTP or inter-process communication (IPC) connection.
This system is more secure than a general web application as the user does not need to give control of her private keys to the service provider, which is not possible in a web application. Moreover, the interaction with the portfolio and the wallet, where the tokens are stored, is administered only by the client-side.

\subsection{Front-end}
As a blockchain works primarily on the client-side, it is essential to have a proper interface for interacting with the Ethereum blockchain directly. Any DApp is similar to a web application in multiple ways. Using various JavaScript web technologies, an easy-to-use interface supports \mech\ Rinkeby blockchain/off-chain interaction. 
MetaMask, that allows us to run Ethereum DApps in the browser without running a full Ethereum node, is used for submitting responses to an SC, to create a response sheet to the SC by the evaluator and other user-friendly interfaces.
Truffle framework is used to create a new wallet and Ganache-CLI is used for Ethereum blockchain simulation. ReactJS (a JavaScript library) is used for building user interfaces especially for the  single page applications such as the enrolment page.  

\subsection{Smart Contract and Ethereum}

An abstract description of the \mech\ SCs is given in \Cref{fig:smartcontract}. Each block in the figure depicts two segments: the first shows the objects needed for the SC, and the second part shows the functions used in the SC.
Using the standard web3.js JavaScript library, \mech\ implements the generic JSON Remote Procedure Call (RPC) specification providing a convenient interface for the RPC methods. The DApp can be accessed through a regular web browser with an extension of MetaMask. The MetaMask connects our DAPP to the main blockchain provided by Ethereum. The MetaMask extension utilizes the web3.js API to an injected web3 object, which can be used to access blockchain using asynchronous methods calls. 

The \mech\ token contract is an {\em ERC-20} token. The ERC-20 token is a standard for implementing SCs on the Ethereum blockchain. The following two sections describe the SCs we use in \mech.

\subsubsection{College}
This segment of contracts contains two contracts, {\em enrolment} and {\em wallet}, that are all sub-standards of the ERC-20 token. In the enrolment contract, various functionalities like sign-up and creation of wallets are implemented. The wallet contract is generated by the enrolment contract. Together, they are analogous to the admission process of an institute, where each student/teacher/evaluator gets an ID card having various functionalities. The whole platform uses \mech\ tokens, which is the currency of the payment service, and allows paying the transaction costs (gas) in these tokens instead of the Ether. The wallet contract has all the functionalities of a usual Ethereum wallet. 
The wallet has the unique function of setting an admin role who can edit the scores (skill and evaluation) and transfer tokens. This role is assigned to the instructor when (s)he is conducting the exam and is withdrawn after the exam is over. The process of assignment and withdrawal is actuated by the enrolment contract.

\subsubsection{Test}
Test segment contains two SCs, {\em exam factory} and {\em exam}, that are called when the examiner wants to add a new exam to the blockchain.  Once the contract is added to the blockchain, functions such as {\tt enrolintest, answermytest} are called by the students to execute similar operations. The $( G^*, \mathbf{r^*}, \mathbf{t^*})$ algorithm is implemented and sends the computed marks to the exam contract, which further distribute scores and rewards to the corresponding users' wallet. The ERC-20 token style, which is followed in \mech, allows the contracts to call the functions of each other. However, these contract also have some private functions which allow certain users to take specific actions, e.g., only the examiner can take the final decisions on the grades.

\section{Summary}

In this paper, we present a blockchain-based education platform \mech, which satisfies scalability, uniformity of scores over multiple evaluators, and ensures honest effort from the evaluators. We presented the detailed design of the platform which is currently active internally in a university network. As a future work, we would like to investigate if such guarantees can be ensured in general peer-review setups.

\appendix

\subsection*{Proof of \Cref{thm:bi-rm}}

According to the model (\S\ref{sec:model}), the instructor may not know the true $y_{j}$ values of each paper $j$. This is because we assume that the student knows her $y_j$ perfectly and if $r_j^* \geqslant y_j$, she does not raise a regrading request.\footnote{It is natural that a student getting a higher score than her expectation will not contest the grading.} Mechanism $\langle G^*, \mathbf{r^*}, \mathbf{t^*} \rangle$ will assume $r_j^*$ to be the true score and design the evaluation scores accordingly when there is no regrading request. The student asks for regrading {\em only if} $r_j^* < y_j$.

To prove \Cref{thm:bi-rm}, we show that the expected value of $t_{i}^j$ is independent of $b_{i}$ and decreasing in $\sigma_{i} := \frac{1}{\tau_i}$.
The final grade after regrading is 
\begin{equation}
 \label{eq:r-diff}
 \max\{r_{j}^{*}(\cdot),y_{j}\}=\max\{r_{j}^{*}(\cdot)-y_{j},0\}+y_{j}.
\end{equation}
We use the shorthand $x = |P_i|, \forall i \in M$ since this number is a constant in our mechanism.

According to the $\mathbf{PG}_1$ model of \citet{piech2013tuned}, $\tilde{y}_j^{(i)}=y_j+b_i+n_{ij}$, where $n_{ij} \sim \mathcal{N}(0,1/\tau_{i})$ is a noise term. Hence, it is easy to show that $\hat{b}_i = b_{i}+\frac{\sum_{k\in P_i} n_{ik}}{x}$ and $\frac{1}{\hat{\tau}_i}=\hat{\sigma}_{i}^{2} = \frac{\sum_{k \in P_i}(n_{ik}-\frac{1}{x}\sum n_{ik})^{2}}{x}$, where $n_{ik} \sim \mathcal{N}(0,1/\tau_{i})$.

Substituting these values we get the expression for
\[r_{j}^{*}(\cdot)-y_{j} = \frac{\sqrt{\gamma}(\mu-y_{j})+\sum_{l\in G(j)}\sqrt{\hat{\tau}_{l}}(n_{lj}-\frac{\sum_{k\in P_l}n_{lk}}{x})}{\sqrt{\gamma}+\sum_{l\in G(j)}\sqrt{\hat{\tau}_{l}}}.\]

With the assumption that $y_j$ is replaced with $r_j^*$ if not regraded, we can find the condition on $y_j$ when the $R(\cdot, \cdot)$ function is non-zero to be $y_{j} \geqslant \frac{\sqrt{\gamma}\mu
+ Z
+\sqrt{\hat{\tau}_{i}}(n_{ij}-\frac{\sum_{k\in P_i}n_{ik}}{x})}{\sqrt{\gamma}}$, where $Z = \sum_{l\in G(j) \setminus \{i\}}\sqrt{\hat{\tau}_{l}}(n_{lj}-\frac{\sum_{k\in P_l}n_{lk}}{x})$.
Note that the RHS of the inequality is independent of $\sigma_i$. To find the expected value of $t_i^{j}$, we need to integrate it w.r.t.\ $y_j$, the limits of which is, therefore, independent of $\sigma_i$.
By definition, the $W_j^{(-i)*}$ component of $t_i^{j}$ is independent of bias and reliability of grader $i$. Hence, we only consider the first component which is dependent on the bias and reliability of grader $i$. We will consider the integral only w.r.t.\ $y_j$ to compute $t_i^j$ and we just showed that the limits of this integral is independent of $\sigma_i$. As $R(r_j^*,y_j)$ is decreasing in $|r_j^*-y_j|$, if we show that $|r_j^*-y_j|$ is independent of $b_{i}$ and increasing in $\sigma_{i}$, then we are done. It can be shown that
\begin{gather*}
\begin{align*}
 \lefteqn{ r_j^*-y_j = } \nonumber \\
 &= \frac{Z_{-i} \sqrt{\frac{\sum_{k\in P_i}(m_{ik}-\frac{1}{x}\sum_{l \in P_i} m_{il})^{2}}{x}} \cdot \sigma_i
 +
  \sigma_i \cdot (m_{j}-\frac{1}{x}\sum_{l \in P_i} m_{il})}
 {X_{-i} \sqrt{\frac{\sum_{k\in P_i}(m_{ik}-\frac{1}{x}\sum_{l \in P_i} m_{il})^{2}}{x}} \cdot \sigma_i +1} 
 \end{align*}
 \label{expr:step3}
\end{gather*}
In the last equality, we substituted $Z_{-i} = [\sqrt{\gamma}(\mu-y_{j})+\sum_{l \in G(j)\setminus \{i\}}\sqrt{\hat{\tau}_{l}}(\tilde{y}_{j}^{(l)}-\hat{b}_{l}-y_{j})]$ and $X_{-i} = (\sqrt{\gamma}+\sum_{l \in G(j)\setminus \{i\}}\sqrt{\hat{\tau}_{l}})$. Also, we substituted $n_{ik} = m_{ik} \cdot \sigma_i$. Since $n_{ik} \sim \mathcal{N}(0,\sigma_{i}^2)$, we get $m_{ik}\sim \mathcal{N}(0,1)$. We see that the absolute value of the above expression is independent of $b_{i}$ and increasing in $\sigma_{i}$.


\bibliographystyle{plainnat}
\bibliography{abb,ultimate,swaprava}

\begin{thebibliography}{17}
\providecommand{\natexlab}[1]{#1}
\providecommand{\url}[1]{\texttt{#1}}
\expandafter\ifx\csname urlstyle\endcsname\relax
  \providecommand{\doi}[1]{doi: #1}\else
  \providecommand{\doi}{doi: \begingroup \urlstyle{rm}\Url}\fi

\bibitem[Alammary et~al.(2019)Alammary, Alhazmi, Almasri, and
  Gillani]{alammary2019blockchain}
Ali Alammary, Samah Alhazmi, Marwah Almasri, and Saira Gillani.
\newblock Blockchain-based applications in education: A systematic review.
\newblock \emph{Applied Sciences}, 9\penalty0 (12):\penalty0 2400, 2019.

\bibitem[Chakraborty et~al.(2020)Chakraborty, Jindal, and
  Nath]{chakraborty2018ensuring}
Anujit Chakraborty, Jatin Jindal, and Swaprava Nath.
\newblock Incentivizing effort and precision in peer grading.
\newblock \emph{arXiv preprint arXiv:1807.11657}, 2020.

\bibitem[Duan et~al.(2017)Duan, Zhong, and Liu]{duan2017education}
Bin Duan, Ying Zhong, and Dayu Liu.
\newblock Education application of blockchain technology: Learning outcome and
  meta-diploma.
\newblock In \emph{2017 IEEE 23rd International Conference on Parallel and
  Distributed Systems (ICPADS)}, pages 814--817. IEEE, 2017.

\bibitem[Farah et~al.(2018)Farah, Vozniuk, Rodr{\'\i}guez-Triana, and
  Gillet]{farah2018blueprint}
Juan~Carlos Farah, Andrii Vozniuk, Mar{\'\i}a~Jes{\'u}s Rodr{\'\i}guez-Triana,
  and Denis Gillet.
\newblock A blueprint for a blockchain-based architecture to power a
  distributed network of tamper-evident learning trace repositories.
\newblock In \emph{2018 IEEE 18th International Conference on Advanced Learning
  Technologies (ICALT)}, pages 218--222. IEEE, 2018.

\bibitem[Funk et~al.(2018)Funk, Riddell, Ankel, and
  Cabrera]{funk2018blockchain}
Eric Funk, Jeff Riddell, Felix Ankel, and Daniel Cabrera.
\newblock Blockchain technology: a data framework to improve validity, trust,
  and accountability of information exchange in health professions education.
\newblock \emph{Academic Medicine}, 93\penalty0 (12):\penalty0 1791--1794,
  2018.

\bibitem[Grech and Camilleri(2017)]{grech2017blockchain}
Alexander Grech and Anthony~F Camilleri.
\newblock Blockchain in education, 2017.

\bibitem[Gresch et~al.(2018)Gresch, Rodrigues, Scheid, Kanhere, and
  Stiller]{gresch2018proposal}
Jerinas Gresch, Bruno Rodrigues, Eder Scheid, Salil~S Kanhere, and Burkhard
  Stiller.
\newblock The proposal of a blockchain-based architecture for transparent
  certificate handling.
\newblock In \emph{International Conference on Business Information Systems},
  pages 185--196. Springer, 2018.

\bibitem[Gupta and Nath(2020)]{Gupta2020}
Jay Gupta and Swaprava Nath.
\newblock Skillcheck: An incentive-based certification system using
  blockchains.
\newblock Technical report, IIT Kanpur, 2020.

\bibitem[Lizcano et~al.(2019)Lizcano, Lara, White, and
  Aljawarneh]{lizcano2019blockchain}
David Lizcano, Juan~A Lara, Bebo White, and Shadi Aljawarneh.
\newblock Blockchain-based approach to create a model of trust in open and
  ubiquitous higher education.
\newblock \emph{Journal of Computing in Higher Education}, pages 1--26, 2019.

\bibitem[Mikroyannidis et~al.(2018)Mikroyannidis, Domingue, Bachler, and
  Quick]{mikroyannidis2018smart}
Alexander Mikroyannidis, John Domingue, Michelle Bachler, and Kevin Quick.
\newblock Smart blockchain badges for data science education.
\newblock In \emph{2018 IEEE Frontiers in Education Conference (FIE)}, pages
  1--5. IEEE, 2018.

\bibitem[Nespor(2019)]{nespor2019cyber}
Jan Nespor.
\newblock Cyber schooling and the accumulation of school time.
\newblock \emph{Pedagogy, Culture \& Society}, 27\penalty0 (3):\penalty0
  325--341, 2019.

\bibitem[Piech et~al.(2013)Piech, Huang, Chen, Do, Ng, and
  Koller]{piech2013tuned}
Chris Piech, Jonathan Huang, Zhenghao Chen, Chuong Do, Andrew Ng, and Daphne
  Koller.
\newblock {Tuned models of peer assessment in MOOCs}.
\newblock \emph{arXiv preprint arXiv:1307.2579}, 2013.

\bibitem[Sharples and Domingue(2016)]{sharples2016blockchain}
Mike Sharples and John Domingue.
\newblock The blockchain and kudos: A distributed system for educational
  record, reputation and reward.
\newblock In \emph{European Conference on Technology Enhanced Learning}, pages
  490--496. Springer, 2016.

\bibitem[Srivastava et~al.(2018)Srivastava, Bhattacharya, Singh, Mathur,
  Prakash, and Pradhan]{srivastava2018distributed}
Abhishek Srivastava, Pronaya Bhattacharya, Arunendra Singh, Atul Mathur,
  Om~Prakash, and Rajeshkumar Pradhan.
\newblock A distributed credit transfer educational framework based on
  blockchain.
\newblock In \emph{2018 Second International Conference on Advances in
  Computing, Control and Communication Technology (IAC3T)}, pages 54--59. IEEE,
  2018.

\bibitem[Williams(2019)]{williams2019does}
Peter Williams.
\newblock Does competency-based education with blockchain signal a new mission
  for universities?
\newblock \emph{Journal of higher education policy and management}, 41\penalty0
  (1):\penalty0 104--117, 2019.

\bibitem[Yuan and Wang(2016)]{yuan2016towards}
Yong Yuan and Fei-Yue Wang.
\newblock Towards blockchain-based intelligent transportation systems.
\newblock In \emph{2016 IEEE 19th International Conference on Intelligent
  Transportation Systems (ITSC)}, pages 2663--2668. IEEE, 2016.

\bibitem[Zhao et~al.(2019)Zhao, Liu, and Ma]{zhao2019design}
Wenshuang Zhao, Kun Liu, and Kun Ma.
\newblock Design of student capability evaluation system merging blockchain
  technology.
\newblock In \emph{Journal of Physics: Conference Series}, volume 1168, page
  032123. IOP Publishing, 2019.

\end{thebibliography}

\end{document}